\documentclass[aps,prd,twocolumn,groupedaddress,nofootinbib,preprintnumbers,superscriptaddress]{revtex4-1}
\usepackage{graphicx,mathtools,tabu,array}
\usepackage{epstopdf}
\usepackage{amsmath}
\usepackage{amsfonts}
\usepackage[colorlinks=true,citecolor=red,linkcolor=blue,breaklinks=true]{hyperref}
\usepackage{accents}
\usepackage[figure, table]{hypcap}
\usepackage{amssymb}
\usepackage{appendix}
\usepackage{comment}
\usepackage{bbold}
\usepackage{xcolor}
\usepackage{slashed}
\usepackage{subfigure}
\usepackage{setspace}
\usepackage{footnote}
\usepackage{multirow}
\usepackage{braket}
\usepackage[normalem]{ulem}
\usepackage[utf8]{inputenc}
\usepackage{mathrsfs}
\usepackage{longtable}
\usepackage{enumitem}

\graphicspath{{Figures/}}

\allowdisplaybreaks

\definecolor{palatd}{RGB}{104, 36, 109}
\definecolor{palatb}{RGB}{0, 56, 168}
\definecolor{palatr}{rgb}{0.745,0.118,0.176}
\newcommand\myshade{80}
\colorlet{mylinkcolor}{palatr}
\colorlet{mycitecolor}{palatb}
\colorlet{myurlcolor}{palatd}

\hypersetup{
  linkcolor  = mylinkcolor!\myshade!black,
  citecolor  = mycitecolor!\myshade!black,
  urlcolor   = myurlcolor!\myshade!black,
  colorlinks = true
}

\begin{document}

\preprint{}

\title{ Neutrinos with refractive masses and the DESI BAO results}

\author{Manibrata Sen}
\email{manibrata@mpi-hd.mpg.de}
\affiliation{Max-Planck-Institut für Kernphysik, Saupfercheckweg 1, 69117 Heidelberg, Germany}
\author{Alexei Y. Smirnov}
\email{smirnov@mpi-hd.mpg.de}
\affiliation{Max-Planck-Institut für Kernphysik, Saupfercheckweg 1, 69117 Heidelberg, Germany}

\begin{abstract}

Due to interactions with dark matter, neutrinos can acquire refractive 
masses which explain the data from oscillation experiments. We study the effects of relic neutrinos with refractive masses on structure formation in the Universe. In the model with a light fermionic mediator, above the resonance energy, $E_R$, associated with the mass of the mediator, refractive masses 
have all the properties identical to the usual vacuum masses.
Below the resonance, refractive masses decrease with neutrino energy, 
however, they cannot be used in the same way as usual masses. We study the dispersion relations and group velocities of such neutrinos and their dependence on redshift. We show that in the epoch of structure formation, relic neutrinos were ultrarelativistic and essentially massless particles for $E_R = (10 - 10^5)\,$eV. This allows to reconcile the values of masses extracted from oscillation experiments with the stringent bounds on sum of neutrino masses from cosmological surveys.

\end{abstract}

\maketitle

\section{Introduction} 
Recent measurements of Baryonic Acoustic Oscillations (BAO) from the first year of operation  
of  the  Dark Energy Spectroscopic Instrument (DESI)~\cite{DESI:2024mwx}, 
along with measurements of the Cosmic Microwave Background (CMB) polarization, temperature and lensing  
by PLANCK~\cite{Planck:2019nip}, as well as data from the ACT~\cite{ACT:2023dou}, 
constrain the sum of neutrino masses 
$\sum m_\nu^{\rm cosm} < 0.072\,{\rm eV}$ at $95\%$ confidence level (C.L). 
This bound disfavors the inverted mass ordering (IO) of neutrinos at $3\sigma$.   
Furthermore, the posterior distribution of $\sum m_\nu^{\rm cosm}$ peaks at zero, 
thus indicating a slight tension with even normal mass ordering (NO). 
A more stringent bound, $\sum m_\nu < 0.043\,$eV at 
$2\sigma$ C.L., was obtained by combining DESI BAO and CMB results with   
Supernova Ia,  GRB  and X-ray observations~\cite{Wang:2024hen}.

Recall that oscillation data being sensitive to neutrino mass-squared differences allows one  
to get a lower bound on neutrino masses. In the case of normal mass ordering (NO), 
$m_3 \geq \sqrt{\Delta m_{31}^2}$ and $m_2 \geq \sqrt{\Delta m_{21}^2}$. 
This leads to a lower bound on the sum of masses: 
$\sum m_\nu^{\rm osc} \approx 0.057\,$eV. 
In the case of inverted ordering (IO), $m_1 \approx m_2 \geq \sqrt{\Delta m_{31}^2}$, we obtain  
$\sum m_\nu^{\rm osc} = 0.11\,$eV. Thus, 
\begin{equation}
\sum m_\nu^{\rm osc} > \sum m_\nu^{\rm cosm} 
\label{eq:sum}
\end{equation}
for IO and even for NO in the peak of posterior distribution.  
Future observations can further strengthen the cosmological bound. 
The reason of this tension could be some cosmological uncertainties. 
In Ref.\,\cite{Craig:2024tky}, it was claimed that the DESI measurement 
seems to favor $\sum m_\nu <0 $, if a positive $\sum m_\nu$ is not imposed {\it a priori}. 
Such a measurement, if confirmed, can point to enhanced matter clustering, 
and can be explained in a scenario where dark matter interacts with itself through long-range forces.

Does the inequality in Eq.\,(\ref{eq:sum}) make sense? Can the sum of neutrino masses determined 
in oscillation experiments be reconciled with the strong cosmological bound? 
Notice that the oscillation masses are determined at present 
epoch, $z = 0$,  at neutrino energies $E > 0.1\,$MeV, and in the solar system. 
In contrast, the cosmological bound concerns the average neutrino mass 
in the Universe, during the epoch of structure formation, $z = 10 - 1000$ 
and at low neutrino energies, $E =  (0.01- 1)\,$eV.
Can this difference explain the inequality in Eq.\,(\ref{eq:sum})? 
Of course, usual neutrino masses,generated by the Higgs vacuum expectation value (VEV) in local interactions, should not change with $E$, $z$ and location. 

There are several possibilities which could explain the inequality in Eq.\,(\ref{eq:sum}). Neutrinos can be massless in the early Universe 
and pick up a mass later due to a late phase transition~\cite{Dvali:2016uhn,Lorenz:2018fzb,Lorenz:2021alz}. 
Neutrino masses can be time-varying, either as a result of 
coupling directly with ultralight scalar fields, or through mixing with sterile neutrinos, 
which in turn, interact with an ultralight scalar~\cite{Fardon:2003eh}. 
In these scenarios, neutrino masses are small in the early Universe, 
but later, after neutrinos become non-relativistic, the masses grow to the present values. 
Other ideas include neutrino decays or annihilation  
to lighter particles, thereby reducing the energy density in massive neutrinos, 
or cooling the neutrinos,  through new interactions with dark matter~\cite{Craig:2024tky}. 

Thus, the cosmological surveys raise an important question 
on the nature of the neutrino mass:  
is it generated through a VEV in a short range interaction, 
or does it have a dynamical or refractive origin? 

In this paper, we explore the possibility that the cosmological limit 
on the sum of neutrino masses can be stronger than that from oscillations. 
We assume  that neutrino masses have a dynamical nature,  
such that  the effective masses of the relic neutrinos were smaller in the epoch of structure 
formation in the Universe than now. 
Scenarios in which neutrinos acquire masses partially or completely through interactions with light 
Dark Matter have been considered 
in~\cite{Berlin:2016woy, Krnjaic:2017zlz, Brdar:2017kbt, Capozzi:2018bps, Choi:2019zxy, Dev:2020kgz, Choi:2020ydp,
Losada:2021bxx, Huang:2021kam, Chun:2021ief, Dev:2022bae, Huang:2022wmz, Davoudiasl:2023uiq, Losada:2023zap, Gherghetta:2023myo}.  

We use the scenario studied in~\cite{Sen:2023uga}, where 
neutrinos acquire an effective mass-squared via scattering 
on ultralight Dark matter (ULDM), mediated by a light fermion. 
We consider the evolution of relic neutrinos with refractive masses and show 
that in the epoch of structure formation, they appear essentially as massless particles,   
thus allowing to satisfy the cosmological bound and the inequality in~Eq.\,(\ref{eq:sum}). 
In Sec.\,\ref{sec:NuPot}, we consider the neutrino potential generated by interactions with DM 
and its dependence on the redshift $z$, the resonance energy and the C-asymmetry of the DM. 
In Sec.\,\ref{sec:NuStruct}, we study the key elements relevant for structure formation: the dispersion relations
of neutrinos and antineutrinos,  the group velocities of relic neutrinos and their evolution, the 
relativistic to non-relativistic transition.  
In Sec.\,\ref{sec:Conc}, we discuss the obtained results and conclude.

\section{Neutrino potential and refractive mass and their evolution} 
\label{sec:NuPot}
We assume that massless neutrinos interact with ULDM composed of  scalar bosons $\phi$, 
via a light fermionic mediator $\chi$, through the effective 
Yukawa interaction $g\, \bar{\chi}\, \nu\, \phi + {\rm h.c}\,$. 
Elastic forward scattering of neutrinos on the cold gas of $\phi$ produces an effective potential~\cite{Sen:2023uga} 
\begin{equation}
    V = \frac{m^2_{\rm asy}}{2 E_R}~ \frac{y - \epsilon}{y^2 - 1}\,\, , 
\, \, \, \, \,\,  \, m^2_{\rm asy} \equiv 
\frac{g^2\,(n_\phi + \bar{n}_\phi)}{m_\phi}\,.
\label{eq:PotPhi}
\end{equation}
Here $m^2_{\rm asy}$ is the effective mass-squared,  
$n_\phi ( \bar{n}_\phi)$ is the number density of ULDM particles  (antiparticles) and $m_\phi$ is the ULDM mass. We define  
$y \equiv E_\nu/E_R$. The resonance energy $E_R = m_\chi^2/(2m_\phi)$, where $m_\chi$ is 
the mass of the mediator and 
$\epsilon \equiv (n_\phi - \bar{n}_\phi)/(n_\phi + \bar{n}_\phi)$ is the C-asymmetry of DM. 
For non-relativistic DM,  the energy density equals $\rho_\phi \simeq m_\phi (n_\phi +  \bar{n}_\phi)$, 
and hence $m^2_{\rm asy}=g^2 \rho_\phi/m_\phi^2$. 
The potential for antineutrinos is given by the same expression as Eq.\,(\ref{eq:PotPhi}) with the
substitution $\epsilon \rightarrow - \epsilon$.

At energies much above the resonance, $y \gg 1$,  the potential becomes 
$V \approx m^2_{\rm asy}/2 E$,  thus reproducing  
the standard term due to neutrino mass in the Hamiltonian of flavor evolution of neutrinos.  
Therefore,  $m^2_{\rm asy}$ 
can be identified with the mass-squared extracted 
from the oscillation experiments,  e.g.,  $m^2_{\rm asy} \approx \Delta m^2_{31}$, {\it etc.}  
There are two  viable regions of parameters of the model 
in which the observed values of oscillation parameters can be reproduced 
and all the existing bounds are satisfied~\cite{Sen:2023uga}: 
(i) $m_\phi \sim  10^{-10}~$eV, $g \sim 10^{-10}$, 
$m_\chi \sim  10^{-4}~$eV, and (ii) 
$m_\phi \sim  10^{-22}~$eV, $g \sim 10^{-21}$, 
$m_\chi \sim  10^{-10}~$eV.

In the opposite case, $y \ll 1$, relevant for the relic neutrinos, 
we have  
\begin{equation}
V \approx \frac{m^2_{\rm asy}}{2 E_R}~\epsilon = 
\frac{g^2}{m_\chi^2} (n_\phi + \bar{n}_\phi) \epsilon
= \frac{g^2}{m_\chi^2} (n_\phi - \bar{n}_\phi) \,,
    \label{eq:Potlow}
\end{equation}
which coincides with the Wolfenstein potential~\cite{Wolfenstein:1977ue}.

Let us consider the evolution of the average potential in the Universe 
for relic neutrinos (see Fig.\,\ref{fig:DispRel}). 
The potential is proportional to the number density of dark matter 
and therefore it increases with $z$ as $(1 + z)^3$:  
\begin{equation}
V(z) = \frac{m^2_{\rm asy}}{2 E_R} \xi (1 + z)^3  ~ \frac{y(z) - \epsilon}{y(z)^2 - 1}\, . 
    \label{eq:PotPhi}
\end{equation}
Here  $y(z) = y_0 (1 + z)$,  $y_0 = E_0/E_R$ and   
$E_0 = 5 \cdot 10^{-4}\,$eV  is the present ($z = 0$) 
average energy of relic neutrinos.  
In Eq.\,(\ref{eq:PotPhi}), $\xi^{-1} \simeq 10^5$  is the local overdensity of DM. 
For $\epsilon \neq 0$ and $y \ll 1$, the expression in Eq.\,(\ref{eq:PotPhi}) simplifies as 
\begin{equation}
V(z) \approx \frac{m^2_{\rm asy}}{2 E_R}\epsilon \xi (1 + z)^3,      
    \label{eq:Potavz}
\end{equation}
that is, it does not depend on the energy of the neutrino. 
In the case of $\epsilon = 0$, we obtain from Eq.\,(\ref{eq:PotPhi}),
\begin{equation}
    V(z) = - \frac{m^2_{\rm asy}}{2 E_R} y_0 ~ \xi (1 + z)^4. 
    \label{eq:Potavz}
\end{equation}
The moduli of this potential increases with $z$ faster than that in Eq.\,(\ref{eq:PotPhi}), 
but has much smaller value at $z = 0$ (see Fig.\,\ref{fig:DispRel}).

We can introduce the refractive mass-squared as 
$\tilde{m}^2 \equiv 2 y V E_R = 2V E$. 
It is proportional to the DM density 
and hence depends on space-time coordinates. For $y\gg 1$, the refractive mass-squared 
has all the properties identical to the vacuum mass-squared. 
Below the resonance,  $y < 1$, 
the mass-squared decreases as $y \epsilon$ for $\epsilon\neq 0$  
and as  $y^2$   for $\epsilon=0$, \emph{i.e.}, a sharper decline. 
Furthermore, the effective mass-squared has opposite
signs for neutrinos and antineutrinos.  
This decrease of modulus of the potential indicates that the bound on sum of usual neutrino masses can be avoided.   
However, as we will see later,  below resonance,  the refractive mass cannot be used 
in the same way as the usual mass, 
and, in particular, it cannot be used in consideration of cosmological bound. 

The decrease of mass with energy also affects kinematical measurements of neutrino mass  in the beta decay, in particular,  
of $^3H$ as used by the KATRIN collaboration~\cite{Aker:2024drp}. 
The decay rate depends on the neutrino mass  
via the neutrino momentum $p_\nu$: $\Gamma \propto p_\nu$.  In the case of the usual mass,
we have  $\sqrt{E_\nu^2 - m_\nu^2} =
\sqrt{(Q_e -T_e)^2 - m_\nu^2}$, where $Q_e$ is the end point of spectrum 
and $T_e$ is the kinetic energy of electron.
In the refractive mass case, $p_\nu = Q_e -T_e - V$.
This would be equivalent to $m_\nu^2 = 2 p_\nu V$, and the latter can be negative.
For $E_R = 100\,$eV,   $|m_{\rm asy}^2| = \Delta m^2_{21}$
relevant for beta decay and
$p_\nu = 1\,$eV, this gives $|m_\nu| \sim 10^{-3}\,$eV -
much below the KATRIN sensitivity.

\section{Neutrinos  with refractive mass and DESI observations }
\label{sec:NuStruct}

The cosmological limit on the sum of neutrino masses 
rely on a determination of  the amount of matter clustering during structure formation. The information 
about clustering is  obtained through an accurate measurement 
of the CMB lensing power spectra (from PLANCK and ACT) and the net 
amount of non-relativistic matter content from the BAO, recently observed by DESI with high precision.

Let us first outline how neutrinos with usual vacuum mass affect structure formation.  
Neutrino masses determine the scale of free streaming, 
$\lambda _{\rm FS} (t) \sim v_g(t,m)/H(t)$, where $v_g$ is the group velocity and $H$ is the Hubble constant. 
At smaller scales, $\lambda < \lambda _{\rm FS}$, neutrinos do not cluster 
and damp density perturbations of DM and baryons.  
Correspondingly, neutrinos erase perturbations with wavenumbers $k$ bigger than 
free streaming wavenumber, $k_{\rm FS}\sim 1/\lambda_{\rm FS}$.  
Essentially, neutrino masses start affecting the power spectrum of perturbations, $P_m(k)$, 
at $k_{\rm nr}$, which corresponds to a transition to non-relativistic regime, when $p_\nu = m_\nu$.  
The damping effect of the power spectrum is determined by 
$f_\nu \equiv (\rho_\nu + \rho_{\bar{\nu}})/\rho_m$ -- the fraction of energy density in neutrinos  
and antineutrinos in the total matter energy density, $\rho_m$.  In the non-relativistic case, 
$\rho_\nu \simeq m_\nu n_\nu$, where $n_\nu$ ($\bar{n}_\nu$) is the number density of neutrinos (antineutrinos).  
There are two main reasons of suppression of the power spectrum at $k > k_{\rm nr}$:  
(i) since neutrinos do not cluster, their fraction $f_\nu$ should be subtracted from 
the power spectrum, and (ii) neutrinos reduce the rate of growth of perturbation by contributing to expansion rate of the Universe (thus weakening the gravitational clustering)
but not clustering themselves.  As a result,   
a suppression of power spectrum equals  approximately~\cite{Lesgourgues:2012uu}
\begin{equation}
\frac{P_{m_\nu}(k)}{P_{0}(k)} = 1 - 8 f_\nu\,,  
\label{eq:mattsupp}
\end{equation}
where $P_0(k)$ is the spectrum  in the case of massless neutrinos. 
Increase of $f_\nu$ due to increase of neutrino mass leads to stronger suppression of the power spectrum. Hence, to estimate the effect of neutrinos 
with refractive mass on structure formation, we need to find their group velocities,  energy density and explore the relativistic to non-relativistic transition.

\begin{figure}[!t]
\includegraphics[width=0.48\textwidth]{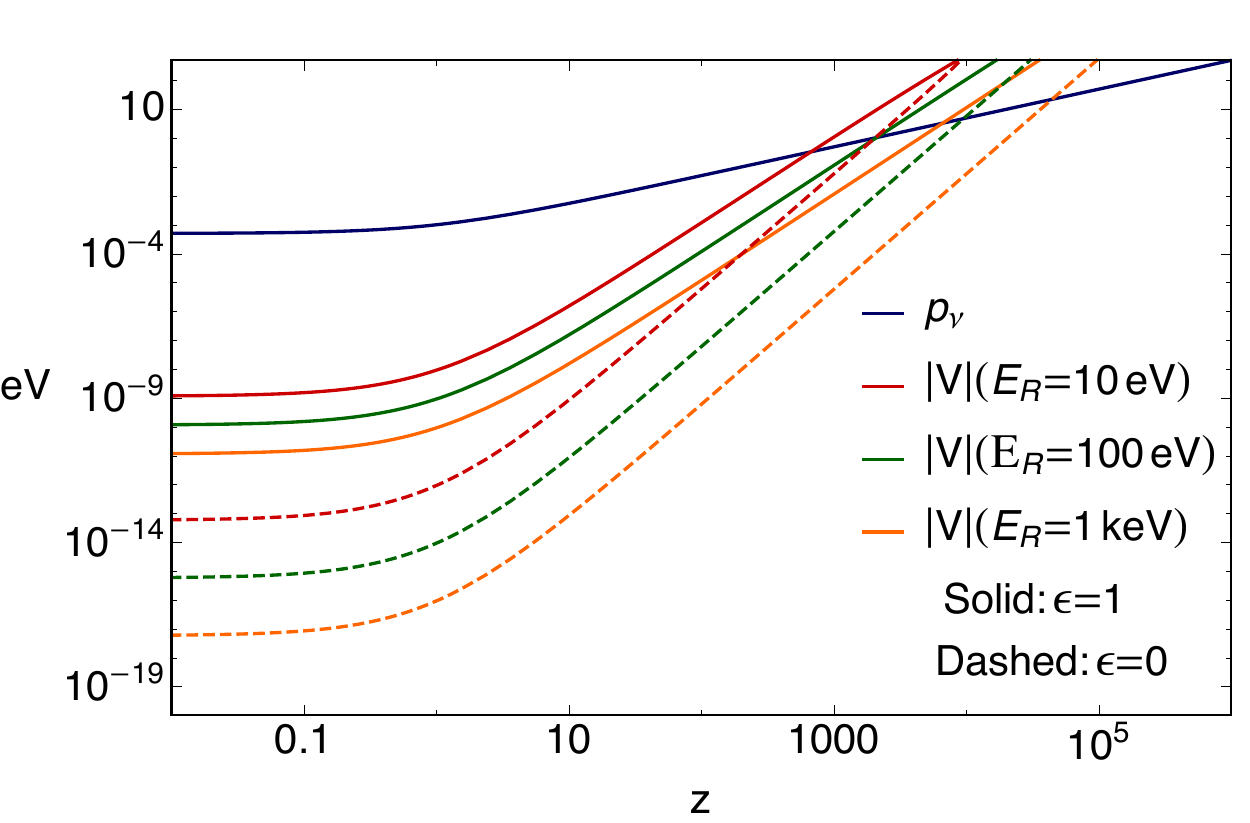}
\caption{Dependence of the neutrino momentum and the effective potential on redshift 
$z$ for different values of the resonant energy $E_R$ and DM charge asymmetry, 
$\epsilon=1$ (solid lines) and  $\epsilon=0$ (dashed lines). We take 
$m_{\rm asy} = \sqrt{\Delta m_{31}^2} = 0.05\,$eV.}
\label{fig:DispRel}
\end{figure} 

For massless neutrinos with refraction effect,  the dispersion relation reads 
\begin{equation}
E = p_\nu + V. 
\label{eq:dispersion}
\end{equation}
Fig.\,\ref{fig:DispRel} shows the dependence of $p_\nu$ 
and $V$ on $z$. For small redshift, one finds $V \ll p_\nu$, 
but as $z$ increases, $V$ increases faster than $p_\nu$ and they become comparable 
at $z \gtrsim 10^3$. An important quantity is the ratio $V/p$ which determines the 
perturbativity of approach~\cite{Sen:2023uga}.
For $y \ll \epsilon \leq 1$, we have
\begin{equation}
\frac{V(z)}{p(z)} = \frac{m_{\rm asy}^2}{2 E_R E_0}
\xi \epsilon (1 + z)^2\,.
\label{eq:voverpE1}
\end{equation}
According to Eq.\,(\ref{eq:voverpE1}), the perturbativity 
condition $V(z)/ p(z) \leq 0.1$  gives 
\begin{equation}
(1 + z) \leq 6.3 \cdot 10^3
\sqrt\frac{E_R}{ (10^2\, {\rm eV})\,\epsilon}\,.
\label{eq:zvpE1}
\end{equation}
For $\epsilon = 0$, we obtain  
\begin{equation}
\frac{V(z)}{p(z)} = - \frac{m_{\rm asy}^2}{2 E_R^2} \xi (1 + z)^3\,,
\label{eq:voverpE0}
\end{equation}
and the inequality  $|V/p| \leq 0.1$ leads to 
\begin{equation}
(1 + z) \leq  3 \cdot 10^4
\left(\frac{E_R}{10^2\,{\rm eV}}\right)^{2/3}\,.
\label{eq:zvpE0}
\end{equation}
For a given $z$, the perturbativity conditions,  
Eqns.~(\ref{eq:zvpE1}) and (\ref{eq:zvpE0}), constrain from below the values of $E_R$.

\begin{figure}[!t]
\includegraphics[width=0.5\textwidth]{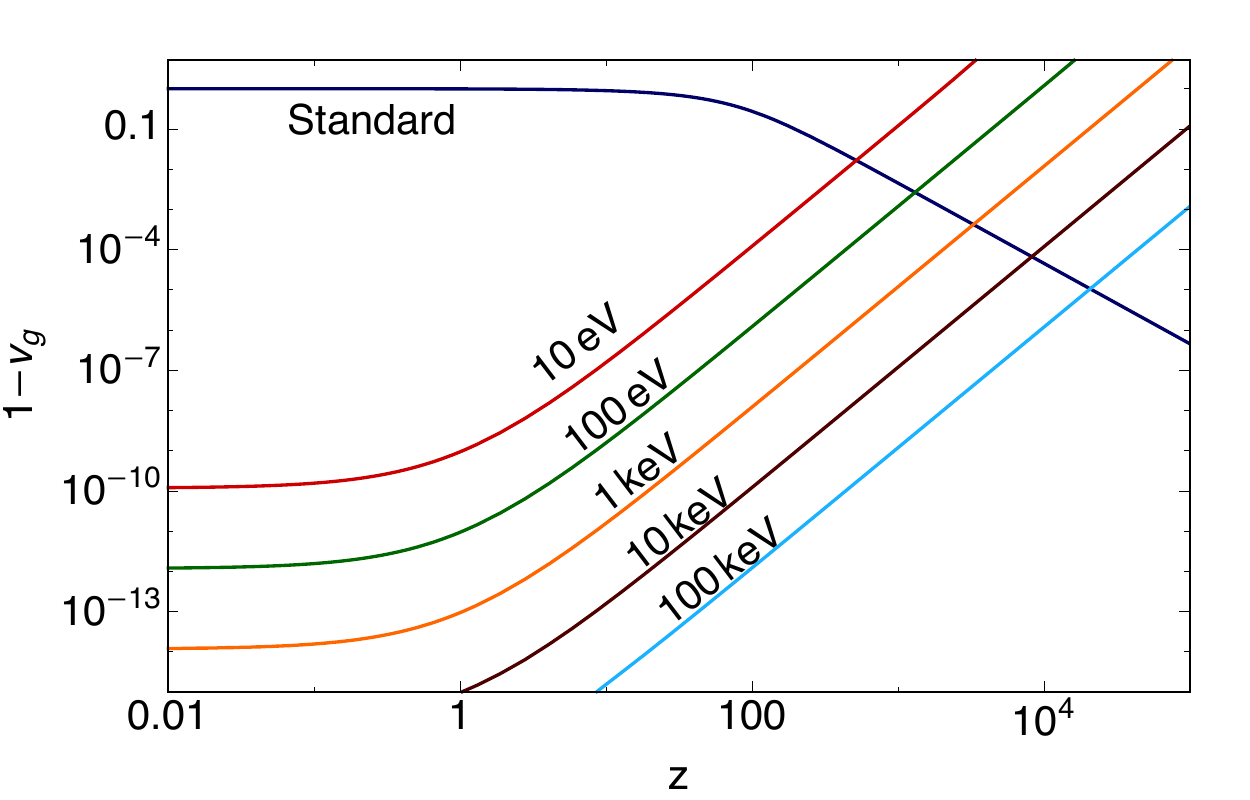}
\caption{The deviation of the group velocity of neutrinos from 1 as a function 
of the redshift for different values of $E_R $ (numberson top of the curves). The lines for $\epsilon=0$ and $\epsilon=1$
practically coincide. 
The standard case for relativistic neutrinos is also shown. We use 
$m_{\rm asy}=  \sqrt{\Delta m_{31}^2} = 0.05\,$eV.} 
\label{fig:GroupVel}
\end{figure}

According to Eq.\,(\ref{eq:dispersion}), 
the group velocity of neutrinos equals  
\begin{equation}
v_{g} = \frac{dE_\nu}{dp_\nu} =  1 - \frac{dV}{dp}\,.
\label{eq:vgroup}
\end{equation}
Using the expression for $V(z)$ from Eq.\,(\ref{eq:PotPhi}),  we find
the deviation of $v_g$ from 1 in the epoch $z$, 
\begin{equation}
1 - v_{g} = 
\frac{\tilde{m}^2_{\rm asy}}{2 E_R^2} \xi (1 + z)^3
\frac{1 + y(z)^2 - 2 \epsilon y(z)}{\left[1- y(z)^2\right]^2}\,.
\label{eq:vgroupz}
\end{equation}
Dependence of the deviation $1 - v_g$ on $z$ is shown in  Fig.\,\ref{fig:GroupVel} for different values of $E_R$.
It is compared with the standard mass case 
$1 - v_g = p/E \approx m^2/2E^2$.   
For small $y$ (realized for parameters in this plot), 
the deviation in Eq.\,(\ref{eq:vgroupz})
does not depend practically  on $\epsilon$,  
in contrast to $V$ and  $\tilde{m}^2$. 
Indeed,  for $y \ll 1$, Eq.\,(\ref{eq:vgroupz}) gives  
\begin{equation}
1 - v_{g} = \frac{\tilde{m}^2_{\rm asy}}{2 E_R^2} \xi (1 + z)^3\,.
\label{eq:vgroupz1}
\end{equation}
The deviation does not depend on energy. Ratio 
of the deviations for refractive and usual masses equals $\xi (1 + z)^3 E^2/E_R^2$, which is small for $E^2 \ll E_R^2$. For 
$m_{\rm asy}^2= m^2 = \Delta m^2_{31}$, they become comparable only for $z>1000$.
According to Fig.\,\ref{fig:GroupVel}, neutrinos with refractive mass $\sim 0.05\,$eV 
are ultrarelativistic in whole time interval of  structure formation upto $z = 10^3$ for $E_R > 10\,$eV. 

Group velocities of neutrinos with usual and refractive masses have an opposite dependence on $z$: in the usual case, with increase of $z$, one has $v_g \rightarrow 1$ and neutrinos become relativistic; on the other hand, 
in the refractive case, $v_g \rightarrow 0$ and neutrinos become non-relativistic. The latter may happen in the epoch of radiation-dominated Universe.

The deviation, Eq.\,(\ref{eq:vgroupz}), differs from the one obtained 
by naive substitution $m^2 \rightarrow \tilde{m}^2$ in the standard formula for the group velocity: 
$1 - v_g = \tilde{m}^2/(2p^2)=V/p$
This is bigger than the estiation in Eq.\,(\ref{eq:vgroupz}) by the factor $\epsilon E_R/p$.  This shows that $\tilde{m}^2$ cannot be used in the same way as usual mass-squared.

The energy densities in neutrinos and antineutrinos equal 
\begin{eqnarray}
\rho_\nu &=& E_\nu n_\nu = (p_\nu + V) n_\nu\,  , 
\nonumber\\
\rho_{\bar{\nu}} &=& E_\nu \bar{n}_\nu = (p_\nu - V) \bar{n}_\nu\, .
    \label{eq:EnDen}
\end{eqnarray}  
Their sum does not  depend on $V$  for  $n_\nu = \bar{n}_\nu$, being the same 
as in the case for massless neutrinos. For non-zero lepton asymmetry, 
$\Delta L \equiv (n_\nu - \bar{n}_\nu)/(n_\nu + \bar{n}_\nu)$, 
we have 
\begin{equation}
\rho_{tot} =  \rho_\nu + \rho_{\bar{\nu}} = p_\nu (n_\nu + \bar{n}_\nu) \left[1 + \frac{V}{p} \Delta L  \right]. 
\label{eq:totalrho}
\end{equation}

\begin{figure}[!t]
\includegraphics[width=0.45\textwidth]{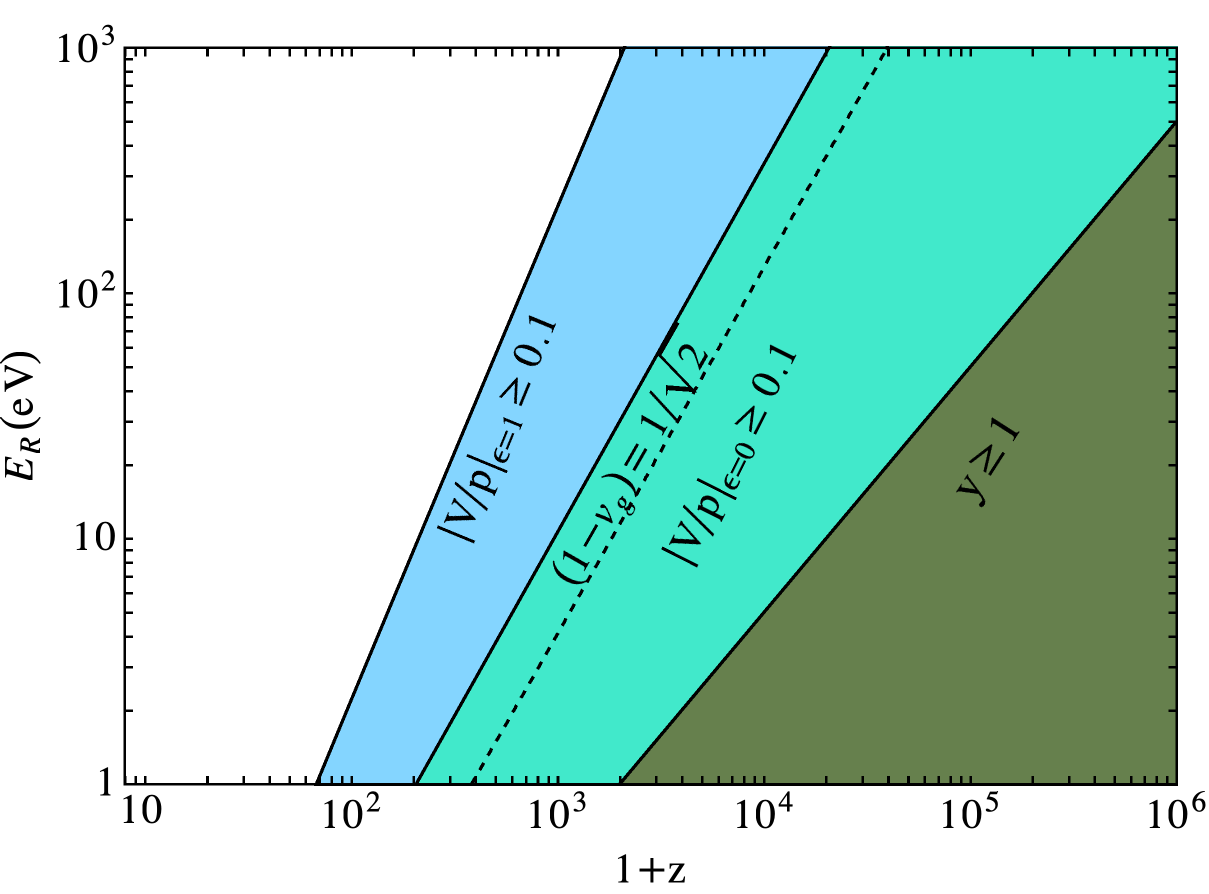}
\caption{Perturbativity constraints $V/p \leq 0.1$ 
in the $E_R - (1+z)$ plane for $\epsilon = 0$ and $\epsilon = 1$. 
The colored parts show  the excluded regions.  
The dashed line corresponds to the group velocity $v_g=1/\sqrt{2}$ or, equivalently to $(1 + z)_{nr}$. 
Also shown is the region of neutrino energies above resonance given by $y\geq 1$.}
\label{fig:const-erz}
\end{figure}

The key parameters relevant for cosmology are $E_R$, $(1 + z)$ and 
$\epsilon$. 
Different bounds in the $E_R - (1 + z)$ plane are shown in Fig.~\ref{fig:const-erz}. 
We present  the lines $V(z)/p(z) = 0.1$  for $\epsilon = 0$ and $\epsilon=1$  
which restrict the region of perturbativity (colored regions are excluded).  
The bounds -- lower on $E_R$ and upper on $(1 + z)$ --  
are stronger in the case of $\epsilon = 1$.  

Interestingly,  for $E_R = (10 - 10^5)\,$eV,  neutrinos with refractive mass are in the perturbation regime at energies relevant for both oscillations  
~\cite{Sen:2023uga} and for relic neutrinos
in epoch of structure formation.

The resonance condition $y = 1$ yields $E_R = E_0 (1 + z)$.
According to Fig.~\ref{fig:const-erz}, the perturbativity requirement is more stringent than $y \leq 1$. So, in the region $V(z) < p(z)$, we have $y \ll 1$, which justifies, in particular, the use of expression in Eq.\,(\ref{eq:vgroupz1}).  

Fig.~\ref{fig:const-erz} also shows the line corresponding 
to $v_g  = 1/\sqrt{2}$, or $p = m$ in usual mass case, that is, the transition to the non-relativistic regime.  
Using Eq.\,(\ref{eq:vgroupz1}), we find an analytical expression of this line: 
\begin{equation}
(1 + z)_{\rm nr} = \left( \frac{E_R^2}{\xi m_{\rm asy}^2} \right)^{1/3}\,.
\label{eq:rwelcond }
\end{equation}
It is shifted by a factor of $5^{1/3}$  to larger values of $(1 + z)$ with respect to the border line of the perturbativity condition for $\epsilon = 0$. 
For bigger values $E_R$ or smaller 
values of $(1 + z)$ with respect to this line, neutrinos can be considered as relativistic. In this region, the situation for structure formation is similar to that for massless neutrinos.  

From Fig.~\ref{fig:const-erz}, for a given $E_R$ at a large $z$, we have  $(1 + z) > (1 + z)_{\rm nr}$, implying
neutrinos are non-relativistic, with $v_g \ll 1$, and consequently, small free-streaming scale. 
This would correspond to $f_\nu \approx 0$ and neutrino contributing to cold DM. As $z$ decreases, $(1 + z) < (1 + z)_{\rm nr}$, neutrinos become relativistic and $f_\nu \simeq f_{\nu} (m = 0)$. For $E_R \sim 1\,$eV,  the transition from non-relativistic to relativistic cases occurs at $z = 10^2 - 10^3$,  thus affecting structure formation. This may appear as some effect of neutrino mass but distortion of the power spectrum is expected to be different.  Note, however, that for $\epsilon = 0$, we have $V/E  = 0.5$ along the border line $(1 + z)_{\rm nr}$, so perturbative consideration may not work well in this range. 

The effective neutrino mass can be defined 
via the group velocity as $m_{\rm eff} = p( v_g^{-2} - 1)^{1/2}$
as in the standard case.  
Using this equation and Eq.~(\ref{eq:vgroupz1}) 
for  $(1 - v_g) \ll 1$, we obtain
\begin{equation}
m_{\rm eff} \approx p \sqrt{2(1 - v_g)} \approx m_{\rm asy} \frac{p}{E_R} 
\sqrt{\xi (1 + z)^3}\,.
\label{eq:meff}
\end{equation}
It should be compared with $\tilde{m} = \sqrt{\tilde{m}^2} = \sqrt{2 pV}$.  
Following Eq.\,(\ref{eq:vgroup}), one can write $m_{\rm eff} = p \sqrt{2 dV/dp} = \sqrt{2 p^2 dV/dp}$. 
Explicitly, the ratio of these masses equals  
\begin{equation}
\frac{m_{\rm eff}}{\tilde{m}} = \sqrt{\frac{p(z)}{E_R}}\,.
\end{equation}
So,  $m_{\rm eff}$ can be orders of magnitude smaller than the  refractive mass $\tilde{m}$. 
This shows again that refractive mass cannot be used at low energies for description of structure formation. 
Correspondingly, the cosmological bound on the parameter space 
of the model is weaker than that obtained in~\cite{Sen:2023uga}. 

\section{Conclusion}
\label{sec:Conc}

We studied the cosmological evolution of neutrinos
with refractive  masses and their influence on structure formation in the Universe.
In the model with a light fermionic mediator, 
below the resonance energy, the refractive masses 
decrease with $E_\nu$, 
indicating that cosmological bound on the sum of neutrino masses can be avoided. 
However, below resonance, one cannot use the refractive masses  
in the same way as the usual (vacuum) masses. Therefore, the effects of neutrinos on structure formation  
should be considered directly without invoking the refractive masses. 

Using the dispersion relations, we computed the group velocity of neutrinos and studied its properties. 
We showed that for $z \leq 10^{3}$ and $E_R > 100\,$eV, 
the deviation $1 - v_g \ll 1$, i.e., 
neutrinos are ultrarelativistic. Furthermore, in this range of $E_R$ and $z$, one finds  $V/p \ll 1$,  
so that both the dispersion relation  
and the total energy practically coincide with those for massless neutrinos.  
Neutrinos with refractive masses appear as massless particles  
in the epoch of structure formation and therefore the cosmological analysis, 
if interpreted in terms of usual neutrino mass, would give zero mass. 
This removes the tension, Eq.\,(\ref{eq:sum}), between the sum of neutrino masses determined from oscillation results and the bounds derived from cosmology. As a result, not only normal ordering and inverted ordering mass spectra but also quasi-degenerate spectrum is allowed.  

Searches of non-zero  neutrino masses in cosmology is a test 
of nature of neutrino masses, and in particular, their possible refraction origins.

\section*{Acknowledgments}

We thank the Galileo Galilei Institute for Theoretical Physics for the hospitality 
and the INFN for partial support during the completion of this work.

\bibliographystyle{JHEP.bst}
\bibliography{biblio.bib}

\end{document}